# Дропшиппинг – новая торговая революция
## М.Л. Калужский

# Dropshipping - a new business revolution
## M.L. Kaluzhsky

*Аннотация*: Статья о новой форме торговли в сети Интернет. Преимущества дропшиппинга открывают для российских производителей и предпринимателей новые горизонты в международной торговле и маркетинге. Автор рассматривает особенности развития и роль дропшиппинга в электронной коммерции нового тысячелетия.

*Ключевые слова*: дропшиппинг, электронная коммерция, интернет-маркетинг, электронная торговля.

*Annotation*: Article about the new form of trade in a network the Internet. Advantages of dropshipping open to the Russian manufacturers and businessmen new horizons in international trade and marketing. The author considers features of evolution and a role of dropshipping in electronic commerce of a new millennium.

Дропшиппинг – сравнительно новое явление в экономической практике, о существовании которого еще 8-10 лет знали только очень узкие специалисты. Да и то, не в России.

Вместе с тем, дропшиппингом можно назвать форму посреднической деятельности, бурно развивавшуюся в России в начале 1990-х гг. на фоне всеобщего дефицита и гиперинфляции. Тем удивительнее, что столь примитивная форма предпринимательства получила неожиданное продолжение в эпоху электронной коммерции. Сегодня она не только привлекает миллионы участников по всему миру, но и на равных конкурирует с экономическими монстрами нового тысячелетия – торговыми сетями и корпоративными каналами распределения.

> **Анекдот начала 1990-х гг.**
> **«Бизнес по-русски»**
> Встречаются два бизнесмена:
> – Тебе вагон повидла нужен?
> – Сколько за него хочешь?
> – Миллион рублей.
> – Согласен, беру!
> Бьют по рукам и расходятся. Один идёт искать миллион рублей, другой – вагон повидла...

Сущность дропшиппинга состоит в том, что дропшиппер, продавая товар от своего имени, на самом деле не обладает ни товаром, ни правами на него. Его бизнес строится на предоставлении услуг поставщикам по сбору заказов на товары, а покупателям – по организации поставок. При этом поставщики не несут никаких затрат по содержанию дропшипперов, которые на свой страх и риск выполняют функции торговых представителей.

**Происхождение дропшиппинга**. Дропшиппинг (*dropshipping*) в переводе с английского означает «прямые поставки». Впервые дропшиппинг был описан в 1927 году американскими маркетологами Х. Мейнардом, В. Вейдлером и Т. Бекманом в учебнике «Принципы маркетинга».[1] Именно поэтому родиной дропшиппинга принято считать США.

Правда следует отметить, что дропшиппинг начала XX века и дропшиппинг начала XXI века – это, как говорят в Одессе, – две большие разницы. Дропшипперы начала XX века торговали углём, коксом, промышленным оборудованием, нефтью, стройматериалами и сельхозпродукцией. Их бизнес был основан на неразвитости коммуникаций. Располагаясь вблизи потенциальных клиентов и обладая информацией о ценах и источниках товара, они без труда находили покупателей.

Отличие дропшипперов от оптовых торговцев заключалось в том, что дропшипперы не имели ни собственных складов, ни сколько-нибудь развитой торговой инфраструктуры. Зачастую их бизнес-инфраструктура ограничивалась столом, стулом, пишущей ма-шинкой и телефоном. Не случайно до сих пор сохранившееся сленговое наименование дропшиппера – «*desk jobber*» – застольный посредник.

---

[1] Maynard H.H., Weidler W.C., Beckman T.N. Principles of marketing. – New York: Ronald Press, 1927.

Нельзя сказать, что поставщики были в восторге от таких посредников. Нередко поставщики наотрез отказывались от сотрудничества с дропшипперами. Основная претензия к ним была связана с неопределённостью сроков отгрузки зарезервированного товара, который поставщик хранил у себя до завершения сделки. Поэтому многим дропшипперам приходилось идти на 100%-ную предоплату товара даже до его перепродажи.

По данным четвертого издания книги Г. Мейнарда, В. Вейдлера и Т. Бекмана «Принципы маркетинга» (1946) в 1939 году на рынке США активно действовало свыше 1000 дропшипперов с продажами 475 млн. долларов, что составляло около 2% от всего объёма оптовой торговли. Среднестатистическая доля трансакционных издержек дропшипперов, обусловленная отказом от функций складирования, погрузочно-разгрузочных работ и т.д., составляла около 6,4% от общего объёма продаж против 21,0% у полнофункциональных оптовых торговцев.[2]

Однако, начиная с конца 1960-х гг., развитие технических коммуникаций постепенно свело на нет преимущества дропшиппинга перед традиционными формами торговли. Во-первых, основные клиенты дропшипперов, промышленные потребители в сфере «B2B», могли самостоятельно делать заказы у поставщиков. Во-вторых, основные конкуренты дропшипперов, оптовые торговцы повсеместно включили методы дропшиппинга в арсенал своих бизнес-инструментов.

В результате дропшиппинг превратился из самостоятельного вида ведения бизнеса в функцию маркетинга. В маркетинговой теории этот переход ознаменовался изданием в 1990 году книги американского маркетолога Николаса Шиля «Дропшиппинг как маркетинговая функция».[3] На этом историю традиционно американского маркетинга в сфере «B2B» можно считать законченной. О дропшиппинге благополучно забыли.

Но ненадолго. Реинкарнация дропшиппинга на новом уровне самоорганизации произошла менее чем через десять лет там же, в США. Только теперь это была уже не сфера «B2B» и традиционная торговля биржевыми товарами, а сфера «B2C» и интернет-торговля потребительскими товарами.

Идея была проста, как всё гениальное: поскольку поставщики создают интернет-сайты и предлагают свои товары для продажи, то почему бы не продавать эти товары с отгрузкой от самих поставщиков? Мало того, можно создать дропшиппера для дропшипперов и брать оплату за доступ к базе данных поставщиков. Также можно привязать к себе поставщиков освободив их от несвойственных им функций по созданию и поддержанию торговых интернет-сайтов.

В интернете не нужен ни офис, ни телефон, а сделки совершаются круглосуточно в автоматическом режиме. Бизнес на дому неминуемо привлекает множество клиентов, превращая их в бесплатных интернет-коммивояжёров поставщиков. Так в 2002 году возникла дропшиппинговая интернет-компания «Doba» – лидер электронного дропшиппинга в США.[4] В качестве иллюстрации можно сказать, что по состоянию на 02.09.2012 г. каталог предлагаемых к продаже товаров этой компании насчитывал 1.447.353 товара в 1.500 категориях (около 8.000 брендов) от 165 поставщиков.

Однако у нового американского дропшиппинга была, есть и будет «ахиллесова пята», не позволяющая США достичь мирового лидерства в дропшиппинге. Эта «ахиллесова пята» – слишком малый дисбаланс («ценовая вилка») между ценой поставщика и розничной ценой в традиционной торговле.

---

[2] Цит. по: Scheel N.T. Drop Shipping as a Marketing Function: A Handbook of Methods and Policies. – Westport (CT): Praeger Publishers, 1990. – P. 4.
[3] Scheel N.T. Drop Shipping as a Marketing Function: A Handbook of Methods and Policies. – Westport (CT): Praeger Publishers, 1990.
[4] Drop Shipping. Simplified.™ [Электронный ресурс] / Сайт дропшиппинговой компании «Doba» (США). – Режим доступа: http://www.doba.com, свободный. – Загл. с экрана. – Яз. англ.

Поэтому подлинное перерождение дропшиппинга оказалось связано не с США, а с Китаем. Именно в Китае дропшиппинг вновь превратился из функции маркетинга в самостоятельный вид предпринимательской деятельности.

Причин такого превращения было несколько. Во-первых, ценовая вилка между розничной ценой и себестоимостью китайских товаров составляла не десятки, а сотни процентов. Во-вторых, электронная коммерция в условиях неразвитости торговой инфраструктуры в Китае является одним из основных методов продаж. В-третьих, только с помощью электронной коммерции китайские товары могли преодолеть сопротивление глобальных торговых сетей и беспрепятственно попасть на мировые рынки.

Так китайские «челноки» и «оптовики» благополучно ушли в интернет. Китайские власти оказали значительную поддержку развитию электронной коммерции, превратив её в инструмент китайской экономической экспансии. Сегодня уровень развития дропшиппинга в КНР является самым передовым в мире.

Предложение китайских поставщиков обеспечивается посредством электронных торговых площадок с многомиллионной аудиторией. Таких, например, как группа компаний «Alibaba Group Holding Ltd», в состав которой входят:

– платёжная система «Alipay» (600 млн. пользователей)[5],
– электронная оптовая торговая площадка («B2B») «Alibaba.com»[6],
– онлайн-рынок розничной торговли «Taobao»[7] с международным подразделением «Aliexpress»[8],
– компания по разработке и продаже готовых решений для управления коммерческой деятельностью в сети Интернет «Alisoft»[9],
– специализированный сайт по обмену Интернет-рекламой между веб-издателями и рекламодателями «Alimama» и др.

Отгрузка покупок из Китая осуществляется посредством логистических компаний, обеспечивающих поставщикам и покупателям экономию до 40% от стандартных почтовых тарифов. В результате только на одном «Alibaba.com» (по данным сайта) сегодня зарегистрировано 72,8 млн. пользователей из 240 стран мира, а объёмы продаж там занимают заметную долю в мировой розничной торговле.

**Первая волна в России: время энтузиастов**. Приход дропшиппинга в Россию неразрывно связан с развитием электронной коммерции. Россия – потребляющая страна. Поэтому дропшиппинг здесь всегда был ориентирован на покупателей, а не на продавцов. Первые дропшипперы были обычными потребителями, обладающими предпринимательской жилкой. Осваивая просторы интернета, они сталкивались с зарубежными формами электронной коммерции и пытались внедрить полученные знания и навыки в российских условиях.

Изначально основным источником поставок была крупнейшая в мире американская торговая интернет-площадка «eBay» (крупнейшая, потому что первая) и китайские торговые площадки: «AliExpress», «DHGate» и целый ряд других. Попадая на эти сайты, российские покупатели с удивлением обнаруживали, что цены на некоторые товары там в разы отличаются от внутрироссийских цен, а ассортимент гораздо шире, чем в розничной торговле.

---

[5] Alipay [Электронный ресурс] / Сайт электронной платежной системы «Alipay». – Режим доступа: http://www.global.alipay.com, свободный. – Загл. с экрана. – Яз. англ.
[6] Alibaba.com [Электронный ресурс] / Сайт торговой площадки «Alibaba.com». – Режим доступа: http://www.alibaba.com, свободный. – Загл. с экрана. – Яз. англ.
[7] Taobao.com [Электронный ресурс] / Сайт онлайн-рынка розничной торговли «Taobao». – Режим доступа: http://www.taobao.com, свободный. – Загл. с экрана. – Яз. кит.
[8] Aliexpress by Alibaba.com [Электронный ресурс] / Сайт онлайн-рынка розничной торговли «Aliexpress». – Режим доступа: http://www.aliexpress.com, свободный. – Загл. с экрана. – Яз. англ.
[9] Alisoft.com [Электронный ресурс] / Сайт компании «Alisoft». – Режим доступа: http://www.alisoft.com, свободный. – Загл. с экрана. – Яз. кит.

После нескольких пробных покупок наиболее предприимчивые потребители решали перейти от дистанционных покупок к дистанционным продажам зарубежных товаров в России. Преимуществом дропшиппинга было то, что он не требовал от дропшипперов ни первоначального капитала, ни складских помещений, ни торгового персонала, ни уплаты налогов и сборов. Любой обладатель компьютера и доступа в интернет мог без всяких вложений пробовать себя на ниве дропшиппинга.

Вместе с тем, были и проблемы, ограничивающие численность потенциальных дропшипперов. Отечественные предприниматели, осваивающие премудрости дропшиппинга, сталкивались с целым рядом трудноразрешимых задач:

1. *Оплата товара*. В России всегда было сложно отладить отправку и получение трансграничных денежных переводов. Международные системы денежных переводов «Western Union» и «Money Gram» взимают слишком высокую плату, делающую их непригодными для оплаты мелких покупок. Система банковских переводов «SWIFT» еще меньшее подходит для этой цели.

Частично проблема решалась с помощью систем денежных переводов «Anelik», «Unistrim» и «Contact». Однако эти системы малоизвестны за рубежом (вне СНГ) и пунктов выдачи денег там явно недостаточно.

Поэтому основным методом осуществления международных расчётов в дропшиппинге было использование международных платёжных систем «PayPal» (США) и «Moneybookers» (Великобритания). Сложность была лишь в том, что многие китайские поставщики отказывались принимать оплату через эти платёжные системы. Их использование было эффективным в основном при оплате сделок поставщикам из Европы и США.

2. *Доставка товаров*. Российская почта, как обычная, так и EMS, всегда славилась непредсказуемостью сроков поставок, вороватостью и безответственностью. Иностранцам часто очень трудно объяснить, почему авиа бандероль из США в Россию доставляется 1,5-2 месяца, тогда как в Европу она идёт не более двух недель. Повлиять на сроки доставки невозможно, равно как и получить законное возмещение за нарушение сроков пересылки, порчу или хищение вложения.

Разумеется, при больших объёмах поставок прибыль от продаж перекрывает неизбежные убытки от работы родной отечественной почты. Страдали и страдают в основном начинающие дропшипперы, для которых необходимость возврата денег даже по одной сделке покупателю может быть критической. Для них Почта России является барьером, сдерживающим приток новых участников этого рынка.

3. *Поиск поставщиков*. Проблема поиска надёжных поставщиков и организации стабильных поставок всегда была наиважнейшей в дистанционной торговле потребительскими товарами. Очень непросто на расстоянии заранее оговорить условия поставок, рекламаций и ответственность поставщика.

Поэтому основные поставки на первом этапе осуществлялись с зарубежных интернет-аукционов и торговых интернет-сайтов крупных западных компаний. Это не всегда было экономически эффективно. Так, на интернет-аукционах действовали посредники, завышавшие цены поставщиков, что снижало прибыль российских дропшипперов. Тогда как интернет-сайты крупных западных компаний весьма неохотно поставляли товары в Россию из-за проблем с оплатой и почтовой доставкой.

Очень немногие энтузиасты могли разобраться в хитросплетении особенностей электронной коммерции и осуществлять эффективную дропшиппинговую торговлю. Развитие этой торговли шло по трём основным направлениям:

1. *Социальные сети*. Дропшипперы заводили аккаунты в социальных сетях и создавали группы, объединяющих любителей специализированных товаров, участников коллективных покупок или просто желающих приобрести оригинальные товары по сниженным ценам. Выкладывая в общий доступ и рекламируя товары, дропшипперы собирали предварительные заявки и после получения аванса заказывали товары в Китае и США.

На первом этапе (2005-2008 гг.) такая форма электронной коммерции привлекала значительное количество покупателей. Отдельные группы социальной сети «Вконтакте» насчитывали до нескольких тысяч активных участников.

2. *Интернет-аукционы*. Торговля на интернет-аукционах подразумевает выставление лота на интерактивные электронные торги, идущие круглосуточно в автоматизированном режиме. Выставление лотов на аукционах в России до недавнего времени было абсолютно бесплатным. При этом пользовательский интерфейс всегда был доступен даже самому неподготовленному пользователю.

Например, крупнейший в России интернет-аукцион «Молоток» не только предоставляет возможность неограниченной торговли товарами, но и компенсирует убытки покупателей на сумму до 5000 рублей в случае недобросовестности продавца. Это делает электронные сделки на «Молотке» гораздо более привлекательными в сравнении с продажами через обычные интернет-магазины.

2. *Интернет-магазины*. Интернет-магазины, несмотря на пропаганду этой формы торговли в сети, обладают весьма ограниченной сферой применения. Во-первых, они требуют наличия специальных технических знаний и усилий по поддержанию их работоспособности. Во-вторых, они требуют специального продвижения в сети, чтобы поисковые системы выдавали информацию о них на первых страницах. В-третьих, степень доверия покупателей к интернет-магазинам невелика.

Поэтому дропшиппинговые интернет-магазины дороги, непопулярны и недостаточно эффективны. Вместе с тем, в интернете можно найти достаточно большое количество узкоспециализированных дропшиппинговых интернет-магазинов. Чаще всего они продают копии дорогих часов известных марок, китайские гаджеты и телефоны. Преимущества этой формы дропшиппинговой торговли перед другими заключается в крайне низкой доле ответственности продавцов перед покупателями. Покупателям предъявлять претензии в случае мошенничества продавца просто некуда.

И всё же, несмотря на сложности и трудноразрешимые задачи, дропшиппинг в России стартовал, превратившись пусть пока не в массовое, но в весьма распространённое явление. Опыт, полученный первопроходцами, не только лёг в основу успешной деятельности лидеров этого рынка, но и повлиял на дальнейшее формирование дропшиппинговой торговли в России.

**Вторая волна в России: время лихоимцев**. Наиболее предприимчивые дропшипперы не ограничились простой продажей товаров через интернет. Их дропшиппинговый бизнес очень скоро обрёл причудливые формы, мимикрировав под малоизвестные в России зарубежные формы продвижения товаров.

Причиной стало отставание в приобретении навыков совершения электронных покупок российских интернет-пользователей. Экономический кризис 2008 года негативно сказался на покупательной способности россиян. Осталось в прошлом массовое потребительское кредитование, резко снизились доходы.

Покупатели пошли в интернет в надежде купить по сниженным ценам брендовые товары. При этом они попадали в ласковые руки посредников, извлекающих прибыль из потребительской неосведомлённости. Своеобразная «болезнь роста» затронула все сферы электронной коммерции, но дропшиппинг пострадал от неё более всего.

*Интернет-магазины*. Сложившийся к середине 2000-х гг. стереотип о дешевизне товаров в интернет-магазинах сыграл нас руку недобросовестным продавцам. Это привело к созданию большого количества интернет-магазинов, предлагающих товары не только по схеме дропшиппинга, но и откровенно обманывающих покупателей.

В целом, следует отметить, что эффективно решать проблемы доставки, приёма платежей, информационного обеспечения продаж могут лишь очень крупные интернет-магазины. Интернет-аудитория представляет собой не только огромный рынок покупателей, но и огромный рынок продавцов.

Самостоятельно продвигать интернет-магазин на таком рынке крайне сложно. Зато можно низкими ценами и несбыточными обещаниями качества привлечь доверчивых клиентов. Благо охват схем недобросовестной торговли предостаточно.

*Сервисы коллективных покупок* представляют собой самую безобидную схему введения в заблуждение интернет-покупателей. Суть этой схемы заключается в том, что организатор сервиса через социальные сети собирает в группы желающих коллективно приобрести партию товара по оптовой цене. Схема подразумевает использование групп в социальных сетях и на интернет-форумах, либо их комбинацию с интернет-сайтом.

В любом случае организатор бизнеса изначально закладывает свою прибыль в стоимость закупаемого товара. Обычно участники группы направляются на сайт продавца с более высокими ценами, а заказ делается у поставщика с более низкими ценами. Главные условия здесь – невозможность сравнить цены различных поставщиков и доверие к организатору сервиса.

Сервисы коллективных покупок особенно популярны при приобретении детских товаров. С одной стороны это вызвано ограниченностью финансовых возможностей молодых семей, а с другой стороны – наличием свободного времени у потенциальных покупателей. Кроме того, эксклюзивность и богатый ассортимент многих таких товаров в интернете также привлекает большое число потенциальных покупателей.

*Псевдо-шоурумы*. Изначально под шоурумом понималось выставочное пространство, оформленное для представления товаров определённой компании (от англ. «Showroom» – демонстрационный зал). Классическими шоурумами можно считать, например, показ новой коллекции дизайнера одежды или выставочный зал в автомобильном салоне.

В российской электронной коммерции шоурумы приобрели подчас весьма причудливые формы, имеющие мало общего с традиционными шоурумами. Такие псевдо-шоурумы представляют собой комбинацию сайта или группы в социальной сети с арендованным помещением, куда посетители могут попасть только по рекомендации или после регистрации на сайте. Большое значение имеет атмосфера таинственности и эксклюзивности, отделяющая товары в шоурумах от товаров в обычной торговле.

Поставляют псевдо-шоурумы прежде всего товары из Китая, удовлетворяющие как минимум одному их трёх критериев: копирующие модные бренды, соответствующие текущей моде или более дешёвые, чем в традиционной торговле. По методам продаж такая форма коммерции больше всего напоминает торговлю фарцовщиков во времена СССР.

Например, один достаточно крупный шоурум, лидирующий в поисковых запросах, позиционирует себя как поставщик товаров с китайской торговой интернет-площадки «Taobao». Покупатель, попадая на эту площадку, обнаруживает множество иероглифов, не поддающихся переводу. Убедившись в невозможности самостоятельных покупок, он возвращается на сайт псевдо-шоурума. Вместе с тем, описанные товары можно без труда обнаружить и заказать на англо- и русскоязычных китайских торговых площадках «Sammydress» и «Aliexpress», расплатившись через платёжный сервис «Qiwi» в ближайшей булочной.

В целом на этом этапе развития дропшиппинга в России можно выделить два направления: *постсоветское* и *американизированное*.

*Постсоветское* направление акцентируется на использовании устоявшихся потребительских стереотипов и предрассудков для продвижения товаров. Особенность дропшиппинга «по-советски» заключается в том, что покупатели, не желающие самостоятельно разбираться в тонкостях электронной коммерции, платят дропшипперам своеобразные «штрафы» за свою некомпетентность.

*Американизированное* направление копирует американские методы дропшиппинга, собирая оплату с новообращённых дропшипперов либо за доступ к базам данных, либо за обучение дропшиппингу, либо за предоставление товаров для торговли. Пример для подражания таких сервисов – американская компания «Doba», берущая плату за доступ к ба-зам данных поставщиков. Больших перспектив в России такой подход не может иметь, так как он основан не столько на дропшиппинге, сколько на сборе денег с доверчивых интернет-пользователей.[10]

Эта волна в дропшиппинге не закончилась до сих пор. Однако с ростом проникновения интернета, развитием мирового экономического кризиса и падением потребительского спроса, коммерческая эффективность такого рода бизнеса будет неуклонно снижаться. На первое место постепенно выходят более добросовестные методы дропшиппинга.

В любом случае основным фактором эффективности дропшиппинга является то, насколько реальную выгоду получат дропшиппер и конечный покупатель от совершения сделки. Поэтому любые недобросовестные интернет-сервисы лишь «снимают пену» с нарастающего потока новых форм электронной коммерции, но никак не за счёт определения его содержания.

**Третья волна в России: время профессионалов**. Основным двигателем развития профессиональной дропшиппинговой торговли служат технологические возможности, открывающие новые горизонты в развитии электронных продаж. Алгоритм достаточно прост. Сначала появляются новые возможности и пионеры-энтузиасты, методом проб и ошибок создающие свой бизнес. Затем передовые способы и технологии продаж получают широкое распространение, их коммерческая эффективность снижается, а численность продавцов-новаторов регулируется естественным отбором.

Основная тенденция развития дропшиппинга на современном этапе связана с продолжением становления инфраструктуры электронной коммерции. Ситуация в России здесь мало чем отличается от общемировых тенденций. Вперёд мы не забегаем, но и значительного отставания тоже не наблюдается.

Безусловным лидером в развитии организованной дропшиппинговой торговли сегодня является крупнейшая в России электронная торговая площадка «Молоток.ру».[11] Её преимущество в качестве места осуществления дропшиппинговых продаж обусловлено широким ассортиментом сервисных услуг, оказываемых как покупателям, так и непосредственно дропшипперам.

Это существенно повышает доступность дропшиппинговой торговли для новых участников рынка. В отличие от открытия собственного интернет-магазина, торговля через «Молоток.ру» подразумевает возможность ведения торговли не только без специальных знаний, но и без каких-либо финансовых вложений. На «Молоток.ру» не требуется ни система управления контентом (CMS), ни услуги программистов, ни покупка доменного имени и хостинга. Все эти услуги пользователи получают сразу и абсолютно бесплатно.[12]

Молотку удалось решить главные проблемы, связанные с организацией дропшиппинговой торговли:

1. *Привлечение покупателей*. Централизованное продвижение товаров на профессиональной основе обеспечивает недостижимые для стандартного интернет-магазина показатели посещаемости. Как минимум продавцы получают бесплатную возможность предлагать товары миллионам посетителей торговой площадки 24 часа в сутки 7 дней в неделю 365 дней в году. Как максимум они могут воспользоваться за плату дополнительными опциями продвижения: от выделения лотов в общем списке до их предложения на стартовой странице торговой площадки.

2. *Система отзывов*. Возможность покупателям и продавцам обмениваться отзывами и арбитраж со стороны администрации «Молоток.ру» не только дисциплинирует участников торгов, но и создаёт атмосферу доверия при проведении сделок. На систему отзывов завязаны статусы «супер-продавца» и «VIP-продавца», а также блокирование аккаунтов участников в случае нарушения правил торгов (неоплаты, непоставки и т.д.). Множество вопросов, связанных с доверием покупателя к продавцу снимается просто после просмотра отзывов покупателей по предыдущим сделкам.

---

[10] См., напр.: Двойной Тест-Драйв [Электронный ресурс]. – Режим доступа: http://www.shop.feelluck.ru/testdrive, свободный. – Загл. с экрана. – Яз. рус.
[11] molotok.ru открытая торговая площадка [Электронный ресурс] / Электронная торговая площадка «Молоток». – Режим доступа: http://www.molotok.ru, свободный. – Загл. с экрана. – Яз. рус.
[12] Компания // molotok.ru открытая торговая площадка [Электронный ресурс]. – Режим доступа: http://molotok.ru/country_pages/168/0/shops/index.php#shops3, свободный. – Загл. с экрана. – Яз. рус.

3. *Программа защиты покупателей.* Важнейшим аргументом в пользу приобретения товаров на «Молоток.ру» служит Программа защиты покупателей. Согласно правилам торговой площадки покупатели имеют право на получение компенсации в сумме до 5000 руб. в случае обмана со стороны продавца. Компенсации выплачивается за счёт «Молоток.ру». Такая форма стимулирования покупок недоступна для обычных интернет-магазинов, что также способствует повышению доверия к товарным предложениям на «Молоток.ру».

Все это в целом позволяет перевести предложения о покупке товаров при помощи дропшиппинга из категории спама в категорию серьёзной электронной коммерции. Например, только по данным 2011 года «Молоток.ру» ежемесячно посещало порядка 10 млн. пользователей (в 2009 г. – 3,5 млн.), обеспечивая ежегодный оборот в 2,7 млрд. рублей. Что касается статистики продаж, то 70% совершённых покупок приходится там на мобильные телефоны, компьютерные аксессуары, одежду и товары быта – классические дропшипинговые товары.[13]

Особо следует отметить значительное изменение маркетинговых приоритетов «Молоток.ру» в 2012 году. Если первоначально «Молоток.ру» позиционировался как крупнейший в России интернет-аукцион, а затем как «Мегамолл в интернете», то сегодня это «Открытая торговая площадка». Изменения выразились в повышении платы за сделку с 3% до в среднем 5% (1,9-8%) и введении платы за выставление лотов (10-40 коп./лот).

Причиной стало желание руководителей торговой площадки расчистить её от «мусорных лотов», т.е. подержанных вещей, обладающих крайне низкой ликвидностью и затрудняющих покупателям поиск нужных товаров. Вслед за торговыми площадками «eBay» и «Allegro» «Молоток.ру» сделал ставку на профессиональных торговцев.

Это обстоятельство не только ознаменовало новую эпоху в развитии площадки «Молоток.ру», но и стало индикатором качественного перехода в развитии отечественной электронной коммерции. В рунете сформировалась значительная прослойка частных торговцев, обладающих необходимым опытом и объёмами продаж, которые стали «золотым фондом» клиентской базы «Молоток.ру».

Экономическая целесообразность предопределила трансформацию интернет-аукционов от электронных барахолок к полноценным электронным торговым площадкам, ориентированным на привлечение профессиональных участников. Так, в 2011 году десятая часть от 60 тыс. продавцов обеспечивала торговой площадке «Молоток.ру» около 95% прибыли.[14] В подобных условиях оптимизация продаж была неизбежна.

Процесс трансформации не завершён до сих пор. Для окончательного перехода к профессиональной торговле «Молоток.ру» должен решить три проблемы:

1. *Упрощение платежей.* Оплата товара является основной проблемой покупателя и продавца при завершении сделок. Далеко не все платёжные сервисы позволяют сделать этот процесс простым и удобным. Для широкого развития дропшипинговой торговли необходимо, чтобы пользователи «Молоток.ру» могли отправлять и принимать платежи с минимальной комиссией (2-3%) посредством интернет-банкинга, т.е. не выходя из дома.

Эталоном может считаться платёжная система «PayPal», где платежи совершаются мгновенно, а стоимость услуг не превышает 4% от суммы платежа.[15] К сожалению, из си-стемы «PayPal» вывести деньги без дополнительных потерь невозможно. «Молоток.ру» объявил о планах интеграции с платёжной системой «PayU», но реальных результатов такой интеграции пока нет.

---

[13] По материалам ИАК «SeoPult.TV»: «Молоток.ру»: простукивая стену // SeoPult.TV [Электронный ресурс]. – Режим доступа: http://www.seopult.tv/programs/moneymaking/molotok_ru_prostukivaya_stenu/text, свободный. – Загл. с экрана. – Яз. рус.

[14] Там же.

[15] PayPal [Электронный ресурс] / Сайт платёжной системы «PayPal». – Режим доступа: https://www.paypal.com, свободный. – Загл. с экрана. – Яз. рус.

2. *Освоение регионов*. Развитие электронной коммерции в обеих столицах Российской Федерации по некоторым показателям даже опережает соответствующие показатели ведущих западных стран. Однако глубинка России пока ещё значительно отстаёт.

Например, доля продаж в 2011 году на «Молоток.ру» в Москве, Санкт-Петербурге и Московской области составила около 45%. Тогда как на всю остальную Россию пришлось примерно 55% всех продаж.[16] С одной стороны, это свидетельствует о недостаточности проникновения интернет-коммерции в регионы страны. Однако, с другой стороны, это отражает высокий уровень недоверия потенциальных покупателей такой форме торговли.

3. *Улучшение сервиса*. Сервис в электронной коммерции призван сделать покупки в интернете столь же обыденными и доступными, как и покупки в традиционной торговле. Применительно к дропшиппингу такой сервис связан с удобством, как для продавцов, так и для покупателей.

Для этого руководству «Молоток.ру» потребуется решить проблемы, связанные с выставлением лотов на торги и обработкой информации о сделках. Пока уровень сервиса здесь значительно отстаёт от зарубежных аналогов «eBay» и «Delcampe». Так, например, на «Молоток.ру» неприлично малое (50) количество знаков в описании лота, нет возможности вторичных предложений, непродуманные товарные категории и т.д.

Разумеется, не стоит сводить перспективы развития дропшиппинговой торговли только к использованию электронных торговых площадок. Другие формы электронной коммерции также используются в профессиональном дропшиппинге.

Однако, издержки при создании собственного интернет-магазина «с нуля» несопоставимы с издержками при ведении торговли на действующей торговой площадке. Достаточно сказать, что для торговли на электронных площадках не требуются ни познания в программировании, ни покупка хостинга и доменного имени, ни усилия по SEO-оптимизации контента.[17] Плюс к тому нужно добавить огромную популярность электронных торговых площадок у целевых категорий покупателей.

Из действующих интернет-магазинов в России сегодня, пожалуй, только интернет-магазин «Ozon.ru» и крупные специализированные интернет-магазины могут на равных конкурировать с «Молоток.ру». «Раскрутить» собственный интернет-магазин с нуля, да ещё и так, чтобы он мог конкурировать с продажами на «Молоток.ру» или «eBay», пожалуй, уже нереально. Время ушло.

**Новые горизонты дропшиппинга**. Дальнейшее развитие дропшиппинга в России, судя по всему, будет связано с развитием интернет-сервисов на основе облачных технологий и аккумулированием отдельных функций аутсорсинговыми компаниями. Всё, что можно объединить и унифицировать – будет объединено и унифицировано.

Глобализация и интернетизация значительно упростили жизнь дропшипперов. С каждым днём всё больше функций технического обеспечения дропшиппинговой торговли выполняются независимыми сервисными структурами. В этих условиях проигрывают те торговцы, которые не хотят или не могут по тем или иным причинам поспевать за стремительными изменениями в развитии электронной коммерции.

Дропшиппинг постепенно сводится к решению двух основных маркетинговых задач: поиску поставщиков и формированию товарного предложения. Всё остальное за дропшипперов делают или будут делать в скором времени сервисные компании:

1. *Осуществление платежей*. Уже сегодня платежи в Интернете можно производить практически мгновенно путём нажатия одной кнопки (например, в платёжных системах «Yandex-деньги», «WebMoney» и «PayPal»). Осталось решить последнюю проблему:

---

[16] По материалам ИАК «SeoPult.TV»: «Молоток.ру»: простукивая стену... – Режим доступа: http://www.seopult.tv/programs/moneymaking/molotok_ru_prostukivaya_stenu/text
[17] Говоря по-русски, продвигать в поисковых системах сайт не нужно.

обеспечить столь же мгновенный вывод денег из платёжных систем на банковский счёт получателя. Если же общая комиссия при этом не превысит 2-3% от суммы сделки, то количество покупок в Рунете станет расти в геометрической прогрессии.

2. *Виртуальные витрины.* Дизайн и функциональное наполнение продающих интернет-страниц всегда было проблемой для дропшипперов. Сегодня эта проблема решается либо с помощью платных (бесплатных) систем управления контентом (CMS), либо с помощью готовых решений, предоставляемых электронными торговыми площадками. Причём, если в первом случае цены снижаются, то во втором случае цены растут. Это лучше всего свидетельствует о том, что покупатели предпочитают торговые площадки.

3. *Обработка заказов.* Больное место всех успешных дропшипперов – обработка поступающих заказов и отслеживание поставок. Это не так актуально при небольших объёмах продаж, но в случае коммерческого успеха обработка заказов становится большой проблемой. В России сегодня пока лишь торговая площадка «eBay» предоставляет соответствующие инструменты, которые при всех плюсах всё-таки довольно дорогостоящи.[18] С отслеживанием почтовых отправлений у нас пока тоже хуже, чем в США, где почтовый сервис «USPS.com» не только информирует о месте нахождения почтового отправления, но и бесплатно сообщает по e-mail обо всех его изменениях в автоматическом режиме.

4. *Доставка товара.* Проблемы с доставкой товара вызваны в первую очередь нареканиями в отношении низкого качества услуг ФГУП «Почта России». Здесь и воровство, и порча в процессе пересылки, и тотальная безответственность почтовых работников, и формализм в обработке претензий и многое другое. С другой стороны, в последние годы уже появилось множество альтернатив, заменяющих или дополняющих услуги, оказываемые государственными монопольными сервисами.

Во-первых, речь идёт о логистических сервисах, призванных упростить и удешевить процесс обработки и доставки международной почтовой корреспонденции. Пионером здесь стали китайские логистические сервисы, позволяющие получать экономию на почтовых расходах по доставке товара до 70% от действующих тарифов.[19] В США пользователи электронной торговой площадки «eBay» также могут значительно сэкономить на почтовых расходах при централизованной отправке товара.[20]

Во-вторых, уже сегодня можно наблюдать процесс становления новой формы логистических услуг, связанных с централизованной выдачей товаров, приобретённых с интернет-торговле. Речь идёт о формировании независимых сетей доставки и выдачи товаров, приобретённых в интернет-магазинах. Сегодня в России существует множество вариантов таких сервисов: от собственных сетей доставки (Lamoda, Ozon и др.) до специализированных сетей почтоматов (Logibox, СДЭК и пр.).

В любом случае общий вектор развития электронной коммерции направлен в сторону углубления сервисной специализации. И дропшиппинг тут не является исключением. Учитывая широкую доступность дропшиппинга и бурное развитие соответствующей инфраструктуры, очень скоро мы столкнёмся с ситуацией, когда успешными дропшипперами будут узкие специалисты, предлагающие специализированные товары по низким ценам. Только так они смогут успешно противостоять экспансии крупных торговых компаний, ускоренными темпами осваивающих сегодня рынки электронных продаж.

Вместе с тем, большую роль в этом процессе играют зарубежные производители и крупные посредники, делающие ставку на стимулирование дропшиппинга в качестве одного из важнейших каналов сбыта товаров. В первую очередь это относится к китайским производителям, получающим таким образом прямой выход на мировые рынки, минуя традиционные каналы распределения, контролируемые западными конкурентами. Российские производители в большинстве своём пока не осознают значимости дропшиппинга для построения сбытовых стратегий.

---

[18] eBay Stores: Subscriptions & Fees // eBay [Электронный ресурс] / Сайт торговой площадки «eBay». – Режим доступа: http://www.pages.ebay.com/storefronts/subscriptions.html, свободный. – Загл. с экрана. – Яз. англ.
[19] См. напр.: PFC Express: Logistics solutions [Электронный ресурс] / Сайт логистической компании «PFC Express». – Режим доступа: http://www.parcelfromchina.com, свободный. – Загл. с экрана. – Яз. англ.
[20] Shipping News // eBay [Электронный ресурс] / Сайт торговой площадки «eBay». – Режим доступа: http:// http://www.pages.ebay.com/sellerinformation/shipping/globalshippingprogram.html. – Загл. с экрана. – Яз. англ.

Многое в этой ситуации сегодня зависит от позиции государства. Сдержать развитие дропшиппинга в России вряд ли удастся. Гораздо продуктивнее направить усилия на ускоренное внедрение дропшиппинговой торговли как в практику экономического образования и переподготовки безработных, так и в практику сбытовой деятельности российских производителей.

Государственная поддержка дропшиппинга не принесёт дополнительных бюджетных поступлений. Однако такая поддержка позволит не только снизить социальную напряжённость в условиях развивающегося экономического кризиса, но и существенно расширит рыночные возможности отечественных производителей.

Сегодня дропшиппинг в России – это бурно развивающееся, массовое явление, оказывающее огромное влияние на развитие электронной коммерции. Потенциал дропшиппинга как принципиально нового подхода к организации продаж ещё далеко не исчерпан и вряд ли будет исчерпан в ближайшие годы. Поэтому главная задача сегодня состоит в том, чтобы не упустить новые возможности в организации продаж, предоставляемые дропшиппингом в интернете.